# Turing Machine with Faults, Failures and Recovery

**Alex Vinokur**
Holon, Israel
alexvn@barak-online.net
alex.vinokur@gmail.com

**Abstract**. A Turing machine with faults, failures and recovery (TMF) is described. TMF is (weakly) non-deterministic Turing machine consisting of five semi-infinite tapes (Master Tape, Synchro Tape, Backup Tape, Backup Synchro Tape, User Tape) and four controlling components (Program, Daemon, Apparatus, User). Computational process consists of three phases (Program Phase, Failure Phase, Repair Phase). C++ Simulator of a Turing machine with faults, failures and recovery has been developed.

## 1. Informal Definition
### 1.1. Structure

In contrast to practical situation a regular Turing Machine never fails [1].
Some Turing machine that may fail is described below.
A Turing machine with faults, failures and recovery (TMF) is (weakly) non-deterministic Turing machine consisting of:

- five semi-infinite (to the right) tapes:
  - ✓ Master Tape,
  - ✓ Synchro Tape,
  - ✓ Backup Tape,
  - ✓ Backup Synchro Tape,
  - ✓ User (Tester) Tape;
- four controlling components (sets of rules)
  - ✓ Program,
  - ✓ Daemon,
  - ✓ Apparatus,
  - ✓ User.

Only daemon has non-deterministic behavior.

## 1.2. Tapes, Alphabets

The *Master Tape* corresponds to the tape of regular deterministic Turing machine.
Head of the *Synchro Tape* is synchronized with head of Master Tape.
The *Backup Tape* is used to back Master Tape data up.
The *Backup Synchro Tape* is used to back Synchro Tape data up.
The *User Tape* is used to perform pure/ideal computation (without faults and failures).

The Master Tape, the Backup Tape and the User Tape are using the same (user-defined) alphabet.
The Synchro Tape and the Backup Synchro Tape are using the same special (embedded, user-independent) alphabet.

## 1.3. Controlling Components, States and Rules
### 1.3.1. States

<u>*Program*</u> contains:
- user-defined states (initial, internal and halting states),
- user-required check-point states (indirectly defined by user),
- embedded (user-independent) shutting down state.

Note 1. A user may mark some of user-defined rules as check-points. Check-point states are derived from these rules.
Note 2. Shutting down state differs from user-defined halting state.

<u>*Daemon*</u> contains three embedded (user-independent) states:
- passive,
- active,
- aggressive.

<u>*Apparatus*</u> contains two embedded (user-independent) states:
- normal,
- emergency.

<u>*User*</u> contains two embedded (user-independent) states:
- tracking,
- stabilizing.

### 1.3.2. Rules

There are three sets of rules:
- Daemon's set of non-deterministic rules.
  The set includes only daemon states.
- Common set of deterministic rules
  The set includes states of all controlling components (program, daemon, apparatus, user).
  In fact, this common set consists of two subsets:
  - program rules including
    ✓ user-defined rules,
    ✓ user-required check-point rules;

- outside rules including
  - ✓ deterministic daemon rules,
  - ✓ apparatus rules,
  - ✓ user rules.
- Daemon-defined set of rules (fault rules).

### 1.3.3. Transitions

Each transition step consists of two half-steps (tacts):

- <u>*First tact*</u>. Daemon performs transition according to non-deterministic rule from passive state to {*passive*, *active*, *aggressive*}.
- <u>*Second tact*</u>. One of three kinds of transitions is performed:
  - ➢ *normal* transition, if daemon is in *passive* state,
  - ➢ *fault* transition, if daemon is in *active* state,
  - ➢ *failure* transition, if daemon is in *aggressive* state.

Note. On the second tact daemon always goes into passive state.

### 1.3.4. Faults and failures

The difference between fault and failure is as follows:

- <u>*Fault*</u> transition:
  - ✓ apparatus stay in normal state,
  - ✓ illegal (daemon-defined) program rule is applied, and the program continues computation.
- <u>*Failure*</u> transition:
  - ✓ apparatus go into in emergency state,
  - ✓ program is unable to continue computation.

## 1.4. Computational process

There are three phases of computational process:

- program phase,
- failure phase,
- repair phase.

### 1.4.1. Program phase

Program phase includes 7 stages.

#### 1.4.1.1. Stage#1. Computation proper

If <u>current program state is **halting** program state</u> then **go to** <u>Stage#7</u>.
If <u>current program state is **check-point** program state</u> then **go to** <u>Stage#2</u>.
Otherwise:

- Computation on Master Tape is performed according to the set of program rules;
- Head of Synchro Tape is synchronized with head of Master Tape.

#### 1.4.1.2. Stage#2. **Computation check**

1) ***Rewind*** Master Tape and User Tape to the left side.
2) ***Compare*** data on Master Tape and User Tape.
   If the data are identical
   ☐ then ***go to*** *Stage#3*
   ☐ else ***go to*** *Stage#5*.

#### 1.4.1.3. Stage#3. **Back up**

1) ***Rewind*** Master Tape and Backup Tape to the left side.
2) ***Write*** Master Tape to Backup Tape.
3) ***Rewind*** Synchro Tape and Backup Synchro Tape to the left side.
4) ***Write*** Synchro Tape to Backup Synchro Tape.
5) ***Go to*** Stage#4.

#### 1.4.1.4. Stage#4. **Check of back up**

1) ***Rewind*** Master Tape and Backup Tape to the left side.
2) ***Compare*** data on Master Tape and Backup Tape.
   If the data are not identical
   ☐ then ***go to*** *Stage#3*.
3) ***Rewind*** Master Tape, Synchro Tape and Backup Synchro Tape to the left side.
4) ***Compare*** data on Synchro Tape and Backup Synchro Tape;
   ***Set up*** the head of Master Tape to continue computation.
   If the data are identical
   ☐ then ***go to*** *Stage#1*
   ☐ else ***go to*** *Stage#3*.

#### 1.4.1.5. Stage#5. **Recovery**

1) ***Rewind*** Master Tape and Backup Tape to the left side.
2) ***Write*** Backup Tape to Master Tape.
3) ***Rewind*** Synchro Tape and Backup Synchro Tape to the left side.
4) ***Write*** Backup Synchro Tape to Synchro Tape.
5) ***Go to*** *Stage#6*.

#### 1.4.1.6. Stage#6. **Recovery check**

1) ***Rewind*** Master Tape and Backup Tape to the left side.
2) ***Compare*** data on Master Tape and Backup Tape.
   If the data are not identical
   ☐ then ***go to*** *Stage#3*
3) ***Rewind*** Master Tape, Synchro Tape and Backup Synchro Tape to the left side.
4) ***Compare*** data on Synchro Tape and Backup Synchro Tape;
   ***Set up*** the head of Master Tape to continue computation.
  If th
   e data are identical
   ☐ then ***go to*** *Stage#1*
   ☐ else ***go to*** *Stage#5*.

### 1.4.1.7. Stage#7. Summary check
1) **Rewind** Master Tape and User Tape to the left side.
2) **Compare** data on Master Tape and User Tape.
    If the data are identical
    - ☐ then **go to** *shutting down* state
    - ☐ else **go to** *Stage#5*.

### 1.4.2. Failure phase
1) **Apparatus** *go into* *emergency* state.
2) **Go to** Repair phase

### 1.4.3. Repair phase
1) **User** *goes into* *stabilizing* state.
2) **Apparatus** *go into* *normal* state
   **User** *goes into* *tracking* state

## 2. Formalized definition (in outline)
### 2.1. Basic Turing machine

Let TM = <$Q, T, I, \delta$, !, b, $q_0, q_{fin}$> be a (regular deterministic one-type) Turing machine with semi-infinite tape.

Elements of the 8-tuple mean as follows:
- ✓ **$Q$** is the set of the states;
- ✓ **$T$** is the symbol alphabet;
- ✓ **$I$** is the alphabet of input symbols; $I \subseteq T$;
- ✓ **$\delta$** : $Q \times T \rightarrow Q \times T \times \{L, R, N\}$ is the transition function (rules).
- ✓ **!** is the left-side-marker symbol; ! $\in T$, but ! $\notin I$;
- ✓ **b** is the empty symbol; b $\in T$, but b $\notin I$;
- ✓ **$q_0$** is the initial state; $q_0 \in Q$;
- ✓ **$q_{fin}$** is the halting state; $q_{fin} \in Q$.

## 2.2. Turing machine with faults, failures and recovery

**Definition**. 5-tape expanded Turing machine TMF (with semi-infinite tapes) corresponding to a (regular deterministic one-type) Turing machine TM is 17-tuple

TMF = <$D$, $U$, $A$, $P$, $T$, $I$, $S$, $\delta$, $\theta$, $\gamma$, $\mu$, $\pi$, !, b, $q_0$, $q_{fin}$, $q_{stop}$>.

Elements of the 17-tuple mean as follows:
- ✓ ***D*** = {passive, active, aggressive} is the set of the (embedded) states of the daemon;
- ✓ ***U*** = {tracking, stabilizing} is the set of the (embedded) states of the user;
- ✓ ***A*** = {normal, emergency} is the set of the (embedded) states of the apparatus;
- ✓ ***P*** is the set of the all program states (see below, the next paragraph)
- ✓ ***T*** is from TM and used as the symbol alphabets for Master Tape, Backup Tape and User Tape;
- ✓ ***I*** is from TM;
- ✓ ***S*** = {!, b, +} is the symbol alphabet for Synchro Tape and Backup Synchro Tape;
- ✓ **δ** is from TM;
- ✓ **θ** : $Q \times T \to Q \times T \times \{L, R, N\}$ is the set of the program rules marked (by user) as check-point rules; $\theta \subset \delta$;
- ✓ **γ** : $Q \times T \to Q \times T \times \{L, R, N\}$ is the set of the function of illegal (fault) program rules; $\gamma \not\subset \delta$;
- ✓ **μ** : $D \to$ 'set of subsets of $D$' is the daemon transition function (rules).
- ✓ **π** : $D \times U \times A \times P \times T \times S \times T \times S \times T \to D \times U \times A \times P \times T \times \{L, R, N\} \times S \times \{L, R, N\} \times T \times \{L, R, N\} \times S \times \{L, R, N\} \times T \times \{L, R, N\}$ is the global transition function (rules).

Note. The π transition function depends on the δ transition function.
- ✓ **!** is from TM;
- ✓ **b** is from TM;
- ✓ $q_0$ is from TM;
- ✓ $q_{fin}$ is from TM;
- ✓ $q_{stop}$ is the shutting down state; $q_{stop} \in P$ ($q_{stop}$ != $q_{fin}$).

Note. Obviously, TMF with one Master Tape can be generalized to TMF with several Master Tapes.

### 2.2.1. Program states

The set of the all program states (***P***) consists of the following subsets and states:
- ***Q*** (from TM),
- ***P1*** - the set of the check-points states defined on the basis of information contained in the θ transition function.
- ***P2*** - the set of the computation check states;
- ***P3*** - the set of the backup states;
- ***P4*** - the set of the backup check states;
- ***P5*** - the set of the recovery states;
- ***P6*** - the set of the recovery check states;
- ***P7*** - the set of the summary check states;
- $q_{stop}$ - the shutting down state.

The states contained in the sets *P2-P7* are embedded states. They don't depend on specific basic Turing machine TM.

### 2.2.2. Transition functions (the sets of the rules)
#### 2.2.2.1. μ-function
Daemon transition function μ contains the following rules:
- μ(passive) = {active, passive, aggressive},
- μ(active) = {passive},
- μ(aggressive) = {passive}.

#### 2.2.2.2. π-function
Global transition function **π** contains the following kinds of the rules:
- Stage#1. Rules of *Computation proper*;
- Stage#2. Rules of *Computation check*;
- Stage#3. Rules of *Back up*;
- Stage#4. Rules of *Check of back up*;
- Stage#5. Rules of *Recovery*;
- Stage#6. Rules of *Recovery check*;
- Stage#7. Rules of *Summary check*.

The rules for Stage#1 are based on the δ and θ transition functions.
The rules for Stages#2-#7 are mostly TM-independent, i.e., don't depend on the basic Turing machine TM.

## 3. C++ Simulator
C++ Simulator (Beta version) of a Turing Machine with faults, failures and recovery can be downloaded at:
- http://sourceforge.net/projects/turing-machine/ (SourceForge.net, the world's largest OpenSource software development website);
- http://alexvn.freeservers.com/s1/turing-s.html.

The C++-program simulates a Turing Machine with faults, failures and recovery.
Following input files define the machine:
- metafile,
- description file,
- state file,
- alphabet file,
- transition file,
- input word(s) file(s).

Each row of metafile contains data related to some Turing machine with faults, failures and recovery:
- name of description file,
- number of master tapes,
- name of states file,
- name of alphabet file,
- name of transition file,
- name(s) of input word(s) file(s).

Description file contains verbal description of the machine [optional].

State file contains list of initial, halting and internal user-defined program states.

Alphabet contains list of empty, input and internal symbols.

Each row of transition contains some transition rule;
- ♦ some rules may be marked as check-points;
- ♦ illegal daemon-defined rules (fault rules) may be added.

Each row of input word(s) contains input word for some tape.

**References**
[1] **Alfred V. Aho, John E. Hopcroft, and Jeffrey D. Ullman**, The Design and Analysis of Computer Algorithms, *Addison Wesley*, 1974.